\documentclass[aps,prb,reprint,groupedaddress,nofootinbib]{revtex4-2}

\usepackage{amsmath,amssymb,bm}
\usepackage{graphicx}
\usepackage{xcolor}
\usepackage{hyperref}
\usepackage{ulem}

\hypersetup{
  colorlinks=true,
  linkcolor=blue,
  citecolor=blue,
  urlcolor=blue
}

\def\bs#1{\boldsymbol{#1}}
\begin{document}

\title{Three-Dimensional Topology from Stacking-Controlled Umklapp Scattering in Large-Angle Twisted Graphite}

\author{Juncheng Li$^{2,3,4}$}
\author{Cong Chen$^{1}$}
\email{congchen@buaa.edu.cn}
\author{Wang Yao$^{2,3,4}$}
\email{wangyao@hku.hk}
\affiliation{$^{1}$School of Physics, Beihang University, Beijing 100191, China\\
$^{2}$New Cornerstone Science Laboratory, Department of Physics, University of Hong Kong, Hong Kong, China\\
$^{3}$State Key Laboratory of Optical Quantum Materials, The University of Hong Kong, Hong Kong, China\\
$^{4}$HK Institute of Quantum Science and Technology, University of Hong Kong, Hong Kong, China}

\date{\today}

\begin{abstract}
Large-angle twisted graphene lies beyond the local-stacking description of conventional moir\'e systems: inequivalent rotation centers define distinct commensurate interfaces whose low-energy interlayer hybridization is governed by intervalley Umklapp tunneling.
For three-dimensional twisted graphite assembled from interfaces with different crystalline symmetries, a symmetry-constrained effective model reveals that competition between nonchiral and chiral tunneling produces one-ring and two-ring nodal-line phases and a $C_3$-protected higher-order topological insulator in the presence of sublattice (chiral) symmetry.
The nodal rings carry integer winding numbers, allowing oppositely wound rings to annihilate into the gapped phase.
We further examine how sublattice-symmetry breaking modifies these phases.
Density functional theory (DFT) and atomistic calculations for $21.8^\circ$ twisted graphite identify the equilibrium structure as a higher-order topological insulator, while compression drives it into a topological Weyl semimetal phase.
These findings establish the stacking sequence of symmetry-inequivalent interfaces as a means of engineering band topology in three-dimensional twisted structures.
\end{abstract}

\maketitle

\section{Introduction}

Twisting van der Waals materials offers a versatile route to engineering electronic states through interlayer hybridization~\cite{MacDonald_NatMater_20220,Andrea_NatPhys_2020,Rubio_NatPhys_2021,Andrea_NatRevMat_2021,FZhang_Nature_2022,Xiaodong_Nature_2021,ExcitionTMD_2022}. Research has centered on small twist angles, where moiré-induced band flattening quenches the kinetic energy and enhances interaction effects, enabling landmark observations of correlated insulators and unconventional superconductivity~\cite{caoCorrelatedInsulatorBehaviour2018a,caoUnconventionalSuperconductivityMagicangle2018b}, orbital magnetism~\cite{FMtBG_Science_2019,tschirhart2021imaging}, and integer and fractional Chern insulating states ~\cite{serlin2020intrinsic,caiSignaturesFractionalQuantum2023a,zeng2023thermodynamic,park2023fractionally,xu2023observation}. 
In small-angle twisted systems, the interlayer tunneling is determined by the local stacking registry which varies smoothly over the moiré length scale. 
The low-energy layer hybridization is dominated by intravalley coupling between nearby Dirac cones~\cite{lopesdossantosContinuumModelTwisted2012}, giving rise to significant renormalization of band dispersion near the neutrality point (e.g. formation of flat band near magic angle~\cite{bistritzerMoireBandsTwisted2011}). 
Within this description, shifting the rotation center or applying an interlayer displacement mainly translates the moiré tunneling texture; the global interface symmetry is assumed to remain unchanged~\cite{bistritzerMoireBandsTwisted2011,zouBandStructureTwisted2018}.

Large-angle twisted graphene falls outside this paradigm. 
The Fermi velocity remains nearly unchanged by interlayer tunneling~\cite{luican2011single,trambly2010localization}, while it is the opposite-valley Dirac cones that hybridize through intervalley Umklapp processes~\cite{mele2010commensuration,Umklapp_PRL_2015,Umklapp_WYao_2019}.
The loss of scale separation renders the local approximation inadequate. Consequently, changing the rotation center or interlayer displacement alters the interlayer registry, producing a different global interface symmetry and a different interlayer tunneling matrix - rather than simply translating the same moiré pattern~\cite{shallcrossQuantumInterferenceTwist2008,meleInterlayerCouplingRotationally2012,li2025domainwall}. 
The allowed tunneling splits into a handedness-independent (nonchiral) part and a handedness-dependent (chiral) part that flips sign when handedness (twisting direction) is reversed; the rotation center determines the interface symmetry and how these two parts are mixed.
A three-dimensional chiral stack with a uniform twist angle can therefore host a wide range of electronic structures - depending on the chosen sequence of interface symmetries (rotation centers).
This raises a natural question: can the stacking-controlled interplay between chiral and nonchiral interlayer tunneling be used to engineer 3D band topology of interest in large-angle twisted graphite?

Here we show that this interplay generates a topological phase diagram spanning topological nodal-line semimetal and higher-order topological insulating phases in the presence of sublattice (chiral) symmetry, with the former evolving into Weyl-semimetal phases upon breaking that symmetry in realistic large-angle twisted graphite.
Taking $\theta = 21.8^\circ$ as a representative example, we consider a periodic three-dimensional stack that combines the $D_6$ and $D_3$ symmetry interfaces selected by different rotation centers. 
We first construct a symmetry-constrained effective tight-binding model that preserves sublattice symmetry. Within this model, the competition between the nonchiral and chiral interlayer tunneling channels gives rise to one-ring and two-ring nodal-line phases, with each nodal ring protected by an integer winding number, as well as a gapped higher-order topological insulator protected by $C_3$ symmetry.
We then introduce sublattice-symmetry-breaking perturbations within the effective model and determine how they reshape these topological phases.
Finally, we turn to realistic $21.8^\circ$ twisted graphite. DFT bulk-band calculations, complemented by Slater--Koster tight-binding calculations of the hinge states, identify the equilibrium structure as a higher-order topological insulator. Upon compression, the system undergoes a transition to a topological Weyl semimetal phase. These results establish the stacking-controlled interplay between chiral and nonchiral interlayer tunneling, realized by intervalley Umklapp processes at large-angle twisted interfaces, as a new route to nontrivial three-dimensional band topology.

\section{Effective Tight-Binding Model}
\label{model}

\begin{figure}[t]
  \centering
  \includegraphics[width=\columnwidth]{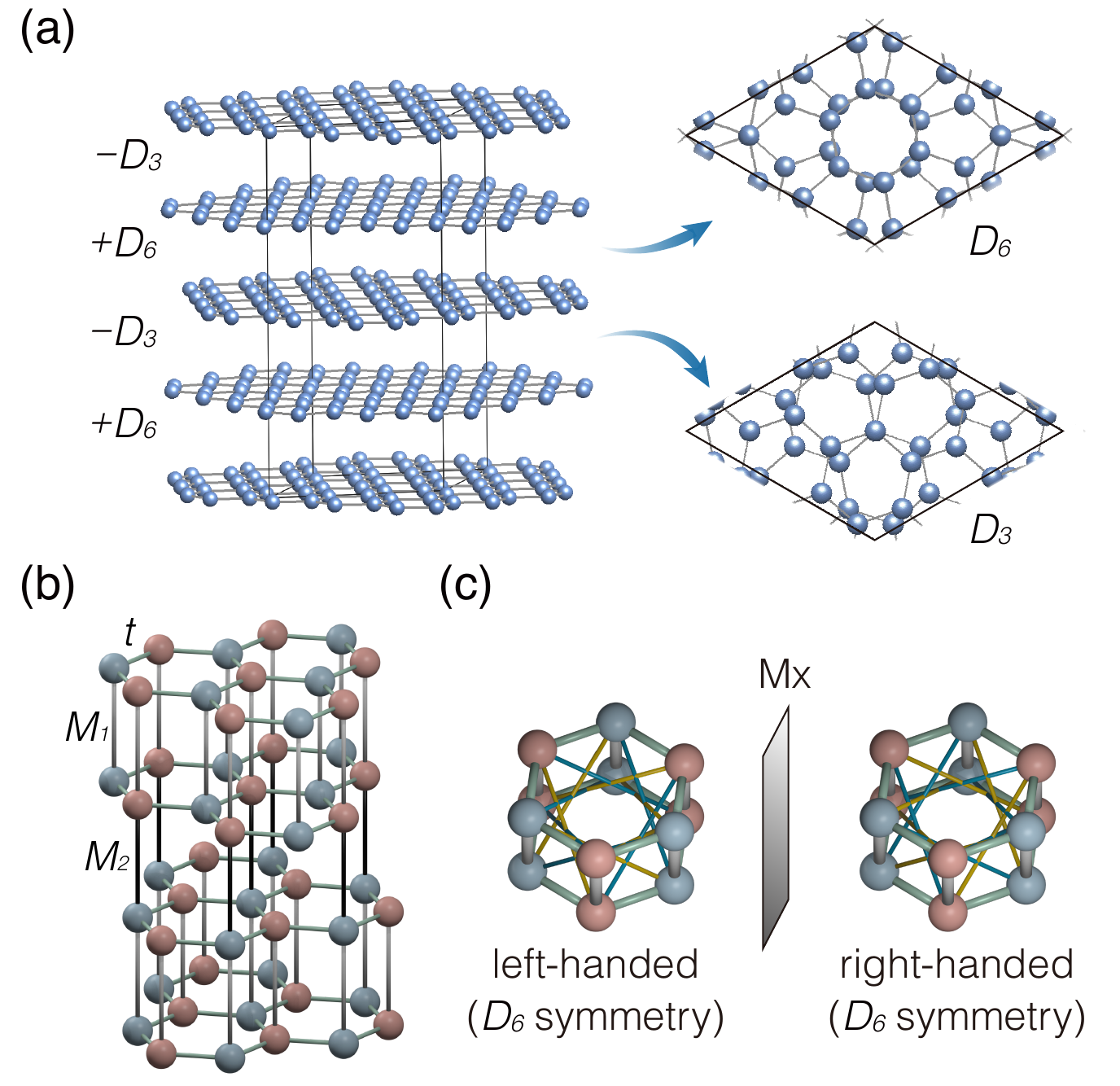}
  \caption{Mixed-symmetry 3D chiral stacking and its effective tight-binding representation in large-angle twisted graphite. (a) Atomic structure of the 3D stacking. The symmetries of the interfaces constitute a periodic sequence of $+D_6$, $-D_3$, $+D_6$, and $-D_3$, where $+$ and $-$ denote opposite twist handednesses. The right panels show top views of the commensurate $\theta=21.8^\circ$ interfaces with $D_6$ and $D_3$ symmetries. A unit cell therefore consists of four atomic layers out-of-plane, and a $\sqrt{7}\times \sqrt{7}$ moire periodicity in-plane. (b) Corresponding effective lattice, in which the $D_6$ and $D_3$ interfaces are represented by AA-like and AB-like interlayer connections, respectively. (c) Mirror-related left- and right-handed bond patterns of the chiral interlayer hopping $\zeta\eta_{ij}\lambda$ on the AA-like interfaces. Dark blue and yellow bonds distinguish opposite signs of the chiral hopping. The mirror operation $M_x$ reverses the structural handedness and hence the sign of the chiral hopping, while leaving the nonchiral hopping unchanged.}
  \label{fig:structure}
\end{figure}

To investigate this stacking-controlled interplay, we consider a three-dimensional periodic structure at $\theta=21.8^\circ$, whose unit cell comprises four twisted atomic layers with a $\sqrt{7}\times\sqrt{7}$ moir\'e periodicity in-plane, as shown in Fig.~\ref{fig:structure}(a). This is made possible with the interface sequence $+D_6$, $-D_3$, $+D_6$, and $-D_3$ along $z$, where $+$ and $-$ denote opposite twist handednesses, providing a periodic chiral platform in which the tunneling channels associated with the two types of chiral interfaces coexist and compete. The $D_6$- and $D_3$-symmetric interfaces are selected by different rotation centers, as shown in the right panels of Fig.~\ref{fig:structure}(a). At this angle, opposite valleys of adjacent layers are folded onto the same moir\'e Brillouin-zone corners and hybridize through intervalley Umklapp tunneling~\cite{zouBandStructureTwisted2018}.

The $D_6$ and $D_3$ interfaces exhibit low-energy dispersions resembling those of renormalized AA- and Bernal (AB)-stacked bilayer graphene, respectively~\cite{mele2010commensuration,moon2013optical}. Despite this spectral resemblance, interlayer tunneling near charge neutrality is dominated by intervalley Umklapp rather than the intravalley coupling of untwisted bilayers~\cite{li2025domainwall}. The symmetry-allowed Umklapp tunneling contains nonchiral and chiral components, which remain invariant and reverse sign, respectively, under reversal of the structural handedness. Accordingly, we represent the $D_6$ interface by an AA-stacked bilayer supplemented by a chiral interlayer hopping, whereas the $D_3$ interface is represented by an AB-stacked bilayer. The much weaker $D_3$ chiral terms are neglected here but included in the material calculations~\cite{chen2025spinless,li2025domainwall}.

The resulting effective lattice consists of four honeycomb layers with alternating AA- and AB-like interlayer connections, as shown in Fig.~\ref{fig:structure}(b). The AA-like connections carry nonchiral hopping $M_1$ and chiral hopping $\zeta\eta_{ij}\lambda$, with the bond-dependent sign pattern shown in Fig.~\ref{fig:structure}(c), while the AB-like connections carry nonchiral hopping $M_2$. The effective unit cell contains eight orbitals, with nearest-neighbor intralayer hopping $t$. Spin-orbit coupling is neglected.

The  spinless effective tight-binding Hamiltonian reads
\begin{align}
H={}&t\sum_{\langle ij\rangle_{\parallel}}c_i^\dagger c_j
+M_1\sum_{\langle ij\rangle_{\mathrm{AA}}}c_i^\dagger c_j
+M_2\sum_{\langle ij\rangle_{\mathrm{AB}}}c_i^\dagger c_j \notag\\
&+\zeta\lambda\sum_{\langle\!\langle ij\rangle\!\rangle_{\perp}}
\eta_{ij}c_i^\dagger c_j+\mathrm{H.c.}
\label{eq:realspace}
\end{align}
Here $c_i^\dagger$ ($c_i$) creates (annihilates) an electron in effective orbital $i$. The sums run over nearest-neighbor intralayer bonds, AA- and AB-like interlayer connections, and next-nearest-neighbor interlayer bonds on the AA-like interfaces, respectively. 
The chiral hopping has magnitude $\lambda\geq0$, with $\eta_{ij}=\pm1$ specifying the bond-dependent sign pattern in Fig.~\ref{fig:structure}(c) and $\zeta=\pm1$ labeling the two opposite structural handednesses. Reversing the handedness sends $\zeta\rightarrow-\zeta$ while leaving $t$, $M_1$, $M_2$, and $\lambda$ unchanged. The explicit Bloch matrix elements are given in Sec. S3 of the Supplemental Material.

The eight orbitals can be partitioned into two effective sublattices such that all retained hoppings connect opposite sublattices, with the partition specified in Eq.~\eqref{eq:sublat_indices} of the Supplemental Material. In the corresponding basis, the Bloch Hamiltonian takes the off-diagonal form
\begin{equation}
H(\mathbf{k})=
\begin{pmatrix}
0&D(\mathbf{k})\\
D^\dagger(\mathbf{k})&0
\end{pmatrix},
\label{eq:offdiag}
\end{equation}
where $D(\mathbf{k})$ is a $4\times4$ matrix. It therefore obeys the sublattice symmetry~\cite{xiaoRevealingSpatialNature2024,zhongPoleZeroEdge2024}
\begin{equation}
\mathcal S H(\mathbf{k})\mathcal S^{-1}=-H(\mathbf{k}),
\qquad
\mathcal S=\operatorname{diag}(I_4,-I_4).
\label{eq:sublattice}
\end{equation}
Together with spinless time-reversal symmetry, $\mathcal T^2=+1$, this places the effective model in symmetry class BDI~\cite{matsuura2013protected}.

\section{Topological Phase Diagram}
\label{phasediagram}

Having established the effective model and its sublattice symmetry, we now determine how the competition between chiral and nonchiral interlayer hoppings reorganizes the bulk spectrum. In the off-diagonal basis, a band crossing at charge neutrality requires~\cite{zhongPoleZeroEdge2024}
\begin{equation}
\det[D(\mathbf{k})]=0.
\label{eq:zero}
\end{equation}
Eq.~\eqref{eq:zero} reduces to
\begin{equation}
\left||f_{1\mathbf{k}}|^2-g_{\mathbf{k}}^2\right|^2
-|g_{\mathbf{k}}|^2M_2^2e^{i\mathbf{k}\cdot\mathbf{a}_3}=0,
\label{eq:det}
\end{equation}
with
\begin{align}
f_{1\mathbf{k}}={}&t\left[1+e^{-i\mathbf{k}\cdot\mathbf{a}_1}
+e^{-i\mathbf{k}\cdot(\mathbf{a}_1+\mathbf{a}_2)}\right],\notag\\
g_{\mathbf{k}}={}&M_1+8i\zeta\lambda
\sin\!\left(\frac{\mathbf{k}\cdot\mathbf{a}_1}{2}\right)
\sin\!\left(\frac{\mathbf{k}\cdot\mathbf{a}_2}{2}\right)\notag\\
&\times\sin\!\left[\frac{\mathbf{k}\cdot(\mathbf{a}_1+\mathbf{a}_2)}{2}\right],
\label{eq:fg}
\end{align}
where $\mathbf{a}_1$ and $\mathbf{a}_2$ are the in-plane primitive lattice vectors, $\mathbf a_3=c\hat{\mathbf z}$ is the out-of-plane lattice vector. Here, we assume $M_1$ and $M_2$ to be positive for simplicity, thus the coefficient $|g_{\mathbf{k}}|^2M_2^2$ is strictly positive. The imaginary part of Eq.~\eqref{eq:det} therefore requires $k_zc=0$ or $\pi$ modulo $2\pi$. At $k_zc=\pi$, the left-hand side is strictly positive, excluding zero-energy solutions. The zero-energy solutions are therefore confined to the $k_z=0$ plane.

In the $k_z=0$ plane, Eq.~\eqref{eq:det} is quadratic in $|f_{1\mathbf{k}}|^2$, with roots
\begin{equation}
|f_{1\mathbf{k}}|^2=M_1^2-n_{\mathbf{k}}^2
\pm\sqrt{M_1^2M_2^2+(M_2^2-4M_1^2)n_{\mathbf{k}}^2},
\label{eq:lattice_roots}
\end{equation}
where
\begin{align}
n_{\mathbf{k}}={}&8\lambda
\sin\!\left(\frac{\mathbf{k}\cdot\mathbf{a}_1}{2}\right)
\sin\!\left(\frac{\mathbf{k}\cdot\mathbf{a}_2}{2}\right)\notag\\
&\times\sin\!\left[\frac{\mathbf{k}\cdot(\mathbf{a}_1+\mathbf{a}_2)}{2}\right].
\label{eq:nk}
\end{align}
Eq.~\eqref{eq:lattice_roots} gives the exact lattice criterion for zero-energy states. For $M_1,M_2,\lambda\ll|t|$, the relevant zero-energy solutions lie near $K$ and $K^\prime$.

We expand the tight-binding model around $K$ in the $k_z=0$ plane and denote the linearized off-diagonal block by $D_K(\mathbf{q})$. After a momentum-independent basis rephasing, the resulting $\mathbf{k}\cdot\mathbf{p}$ Hamiltonian retains the off-diagonal form
\begin{equation}
H_K(\mathbf{q})=
\begin{pmatrix}
0&D_K(\mathbf{q})\\
D_K^\dagger(\mathbf{q})&0
\end{pmatrix},
\label{eq:kp_ham}
\end{equation}
with
\begin{equation}
D_K(\mathbf{q})=
\begin{pmatrix}
vq_-&\mu_-&0&M_2\\
\mu_-&vq_+&0&0\\
0&M_2&vq_+&\mu_+\\
0&0&\mu_+&vq_-
\end{pmatrix}.
\label{eq:kp_d}
\end{equation}
Here $v=\sqrt3t/2$, $\mu_\pm=M_1\pm i\zeta\lambda'$, $q_\pm=q_x\pm iq_y$, $\mathbf q=\mathbf k-\mathbf K$, and $\lambda'=3\sqrt3\lambda$. 

Solving $\det[D_K(\mathbf{q})]=0$ gives
\begin{align}
|\mathbf{q}|^2&=\frac{4}{3t^2}
\left(M_1^2-\lambda'^2\pm\sqrt{\Delta}\right),\notag\\
\Delta&=M_1^2M_2^2+(M_2^2-4M_1^2)\lambda'^2.
\label{eq:kp_roots}
\end{align}
Away from phase boundaries, each distinct positive real solution of Eq.~\eqref{eq:kp_roots} defines a circular nodal ring in the $k_z=0$ plane within the linearized model, with the plus and minus branches giving the outer and inner rings, respectively. The spectrum is gapped when $\Delta<0$ or the larger real root is negative. For $\Delta>0$, one positive and one negative root define the one-ring nodal-line (1R-NL) phase, whereas two positive roots define the two-ring nodal-line (2R-NL) phase. A vanishing root describes a band touching at the valley, while $\Delta=0$ with a positive repeated root describes the merger of the two rings. The continuum solution applies when the outer radius remains small compared with the Brillouin-zone scale, as ensured by $M_1,M_2,\lambda'\ll|t|$. 

\begin{figure}[t]
  \centering
  \includegraphics[width=\columnwidth]{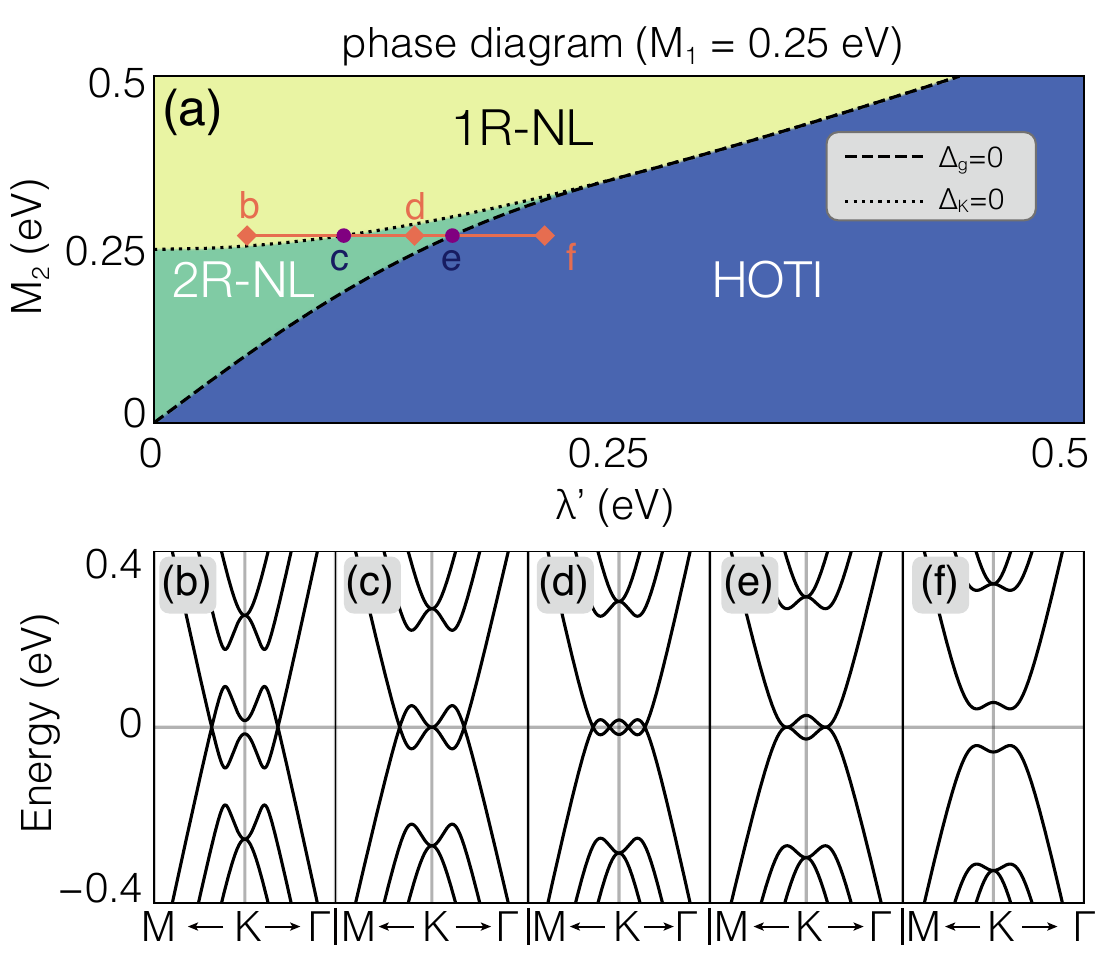}
  \caption{Topological phase diagram and representative bulk-band evolution for fixed $t=-2.66$ eV and $M_1=0.25$ eV. (a) Phase diagram in the $(\lambda',M_2)$ plane, comprising the one-ring nodal-line (1R-NL), two-ring nodal-line (2R-NL), and higher-order topological insulating (HOTI) phases. Here, $\Delta_g$ denotes the global band gap, while $\Delta_K$ denotes the band gap at $K$ in the $k_z=0$ plane. The dashed curve marks the boundary $\Delta_g=0$, across which the global gap opens, whereas the dotted curve marks $\Delta_K=0$. The horizontal cut at $M_2=0.27$ eV follows the sequence $b\to c\to d\to e\to f$. (b)--(f) Low-energy bulk bands along $M$--$K$--$\Gamma$ at points $b$--$f$, corresponding to $\lambda'=0.05$, $0.10$, $0.14$, $0.16$, and $0.21$ eV, respectively. Panels (b), (d), and (f) represent the 1R-NL, 2R-NL, and HOTI phases, respectively, while panels (c) and (e) show the critical spectra at the valley-touching transition and the nodal-ring-merger transition, respectively.}
\label{fig:phase}
\end{figure}

Writing $r=M_1/M_2$ makes the different phase sequences under increasing $\lambda'$ explicit. At $\lambda'=0$, the system has one ring for $r<1$ and two rings for $r>1$. For $r\leq1/\sqrt{2}$, increasing $\lambda'$ drives the 1R-NL phase directly into the gapped phase. For $1/\sqrt{2}<r<1$, an inner ring first appears, producing the sequence 1R-NL, 2R-NL, and gapped. For $r>1$, the initial 2R-NL phase becomes gapped at sufficiently large $\lambda'$. Numerically tracking the zero-energy solutions of the effective  model across parameter space yields the phase diagram in Fig.~\ref{fig:phase}(a), whose phase boundaries are well described by the continuum result in Eq.~\eqref{eq:kp_roots} in the low-energy regime.

To illustrate the phase evolution, we follow the horizontal path $M_2=0.27$ eV in Fig.~\ref{fig:phase}(a), keeping $r$ fixed while increasing $\lambda'$. At point $b$, the system hosts a single nodal ring around $K$. As $\lambda'$ increases, the valley gap closes at point $c$, where the horizontal path intersects the $\Delta_K=0$ curve. Beyond this transition, an inner nodal ring emerges from $K$, yielding two concentric nodal rings at point $d$. Within the continuum description, the inner ring expands while the outer ring contracts until they meet and annihilate at point $e$ on the boundary where the global gap opens. Further increasing $\lambda'$ gives the fully gapped spectrum at point $f$. The representative phases are shown in Figs.~\ref{fig:phase}(b)-(f). Notably, in the solutions of the effective model, the chirality symbol $\zeta$ is eliminated, indicating the invariance of phase diagram when the structural chirality is reversed.

The nodal rings are protected by the chiral symmetry of the BDI Hamiltonian. For a closed momentum-space loop $\mathcal C$ linking a nodal ring, the associated winding number is~\cite{matsuura2013protected}
\begin{equation}
\nu_{\mathcal C}=\frac{1}{2\pi i}\oint_{\mathcal C}
d\mathbf{k}\cdot\nabla_{\mathbf{k}}\ln\det D(\mathbf{k}).
\label{eq:winding}
\end{equation}
For the representative 1R-NL and 2R-NL phases in Figs.~\ref{fig:phase}(b) and \ref{fig:phase}(d), each nodal ring carries a unit winding number, $|\nu_{\mathcal C}|=1$. In the 2R-NL phase, the inner and outer rings carry opposite winding numbers for a common orientation convention of the linking loops and can therefore annihilate when they meet. This integer invariance distinguishes both nodal-line phases from accidental band crossings.

The winding number establishes the topological stability of the nodal rings, but it does not determine whether the gapped phase in Fig.~\ref{fig:phase}(f) is topologically nontrivial. We therefore regard the three-dimensional Hamiltonian as a family of two-dimensional subsystems parameterized by $k_z$. Because the stack preserves $C_3$ symmetry, the gapped subsystems on the time-reversal-invariant planes $k_z=0$ and $k_z=\pi/c$ can be characterized by the rotation indices \cite{benalcazar2019corner},
\begin{equation}
\chi(k_z)=\bigl([K_1],[K_2]\bigr),
\qquad
[K_m]=n_K^{(m)}-n_\Gamma^{(m)},
\label{eq:rotation}
\end{equation}
where $n_{\Pi}^{(m)}$ counts the occupied states at $\Pi=\Gamma,K$ with rotation eigenvalue $e^{2\pi i(m-1)/3}$. For the gapped phase in Fig.~\ref{fig:phase}(f), we obtain $\chi(0)=\chi(\pi/c)=(-2,1)$. The associated vanishing polarization and nonzero $C_3$-quantized corner charge identify the higher-order topology of these two-dimensional subsystems~\cite{benalcazar2019corner,CnHOTI_PRB_2022}. Because the three-dimensional bulk gap remains open as $k_z$ varies, the higher-order topology is extended throughout the whole Brillouin zone.

\section{Sublattice-Symmetry Breaking}
\label{subsymbreak}

The phase diagram above relies on exact sublattice symmetry, which constrains the Bloch Hamiltonian to the off-diagonal form in Eq.~\eqref{eq:offdiag} and pins the nodal rings to zero energy. 
In realistic materials, same-sublattice hoppings and onsite potentials generally break this symmetry. We now examine how these perturbations modify the nodal-line phases and under what conditions the higher-order topological insulating phase survives.

We consider a sublattice-symmetry-breaking perturbation of the form
\begin{equation}
V(\mathbf{k})=
\begin{pmatrix}
A_{4\times4}(\mathbf{k})&0\\
0&B_{4\times4}(\mathbf{k})
\end{pmatrix}.
\label{eq:perturbation}
\end{equation}
Here $A$ and $B$ act within the two sublattice sectors. This perturbation commutes with $\mathcal S$ and thus breaks the sublattice symmetry. Consider a momentum $\mathbf{k}_0$ on a nodal ring of the unperturbed Hamiltonian. The two zero modes satisfy
\begin{equation}
D^\dagger(\mathbf{k}_0)|u_+\rangle=0,
\qquad
D(\mathbf{k}_0)|u_-\rangle=0.
\label{eq}
\end{equation}
Here, $|u_+\rangle$ and $|u_-\rangle$ are even and odd eigenstates of $\mathcal{S}$, respectively. To first order, projecting $V$ onto this zero-mode subspace spanned by $|u_+\rangle$ and $|u_-\rangle$ gives
\begin{equation}
V_{\mathrm{eff}}(\mathbf{k}_0)=
\begin{pmatrix}
V_+(\mathbf{k}_0)&0\\
0&V_-(\mathbf{k}_0)
\end{pmatrix}
=\epsilon(\mathbf{k}_0)I_2+m(\mathbf{k}_0)\sigma_z,
\label{eq}
\end{equation}
where $V_\pm=\langle u_\pm|A|u_\pm\rangle$, $\epsilon=(V_++V_-)/2$, and $m=(V_+-V_-)/2$.

The scalar term $\epsilon I_2$ shifts the two degenerate states equally, whereas the mass term $m\sigma_z$ splits them and opens a local gap of $2|m|$. An in-plane next-nearest-neighbor hopping that contributes a momentum-dependent scalar term $\epsilon(\mathbf{k})I_2$ likewise preserves the degeneracy but reshapes the nodal-line pattern. By contrast, a staggered potential on the two effective sublattices generates a momentum-independent mass term and opens a full gap.

In realistic materials, weak sublattice-symmetry-breaking perturbations generally contain both scalar and mass components, with the latter gapping most of the nodal structure. An additional crystalline symmetry that exchanges the two zero modes can enforce $V_+(\mathbf{k})=V_-(\mathbf{k})$, and hence $m(\mathbf{k})=0$, at symmetry-invariant momenta. Where the deformed band-crossing locus coincides with these symmetry-enforced zeros, isolated point nodes can remain. Their Weyl character is established by nonzero Berry-flux charges, as demonstrated in the material calculations below. The ideal nodal-line phase can therefore evolve into a Weyl semimetal rather than becoming fully gapped.

The gapped phase behaves differently. Its higher-order topology is diagnosed by the $C_3$ rotation indices rather than by the BDI winding number \cite{matsuura2013protected}. 
The gapped HOTI therefore remains stable against weak sublattice-symmetry-breaking perturbations, with $\chi=(-2,1)$ unchanged, as long as the bulk gap remains open and $C_3$ symmetry is preserved. 
In the following, atomistic and DFT calculations examine how these two outcomes are realized in large-angle twisted graphite.

\begin{figure}[t]
  \centering
  \includegraphics[width=\columnwidth]{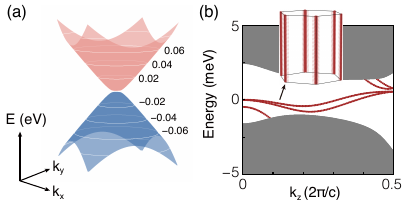}
\caption{Higher-order topology and hinge states in equilibrium $21.8^\circ$ twisted graphite with alternating $D_6$ and $D_3$ interfaces. (a) Low-energy bands near $K$ in the $k_z=0$ plane obtained from DFT calculations, showing a bulk gap of approximately $4$~meV. (b) Slater--Koster tight-binding spectrum of a hexagonal prism finite in the $xy$ plane and periodic along $z$. The gray regions represent bulk and surface bands, while the brown points denote in-gap hinge states forming  bands along the stacking direction. Inset: spatial distribution of the three in-gap states at $k_z=0$, showing localization along the six hinges with a $C_3$-symmetric pattern.}
\label{fig:HOTI}
  \label{fig:HOTI}
\end{figure}

\section{Materials Realization}
\label{mater}

We now turn to the periodic four-layer $21.8^\circ$ twisted structure introduced in Sec.~\ref{model} and restore the atomistic hopping processes omitted from the sublattice-symmetric effective model. 
These calculations include remote same-sublattice hoppings and the much weaker chiral terms of the $D_3$ interfaces omitted from the effective model. 
The results can be understood within the unified theoretical framework developed above.

\subsection{Higher-Order Topological Insulator}

\begin{figure}[t]
  \centering
  \includegraphics[width=\columnwidth]{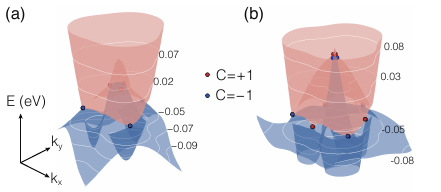}
  \caption{Pressure-induced Weyl phases in the mixed-symmetry $D_6$--$D_3$ stack. (a,b) Low-energy bands near $K$ at hydrostatic pressures of 50 GPa and 62 GPa, respectively. At 50 GPa, three symmetry-related pairs of Weyl points lie along the in-plane $C_2$ axes. Red and blue circles distinguish Weyl points with opposite topological charges, $C=+1$ and $C=-1$, as determined by Berry-flux calculations. At 62 GPa, additional pairs of Weyl points appear.}
  \label{fig:Weyl_semimetal}
\end{figure}

The mixed-symmetry $D_6$--$D_3$ stack in Fig.~\ref{fig:structure}(a) is insulating near charge neutrality, as shown by the DFT band structure in Fig.~\ref{fig:HOTI}(a), with a bulk gap of approximately $4$~meV. For a hexagonal prism finite in the $xy$ plane and periodic along $z$, Slater--Koster tight-binding calculations yield the spectrum in Fig.~\ref{fig:HOTI}(b), where in-gap states form bands as a function of $k_z$. Their spatial distribution, shown in the inset of Fig.~\ref{fig:HOTI}(b), reveals localization along the six hinges of the prism. Together, the bulk gap and hinge-localized states support the realization of the $C_3$-protected HOTI of the effective model even though exact sublattice symmetry is absent, consistent with the stability analysis in Sec.~\ref{subsymbreak}. 
Details of the density-functional and Slater-Koster tight-binding
calculations are provided in Secs. S1 and S2 of the Supplemental Material, respectively.

\subsection{Pressure-Induced Weyl Semimetal}

Compression modifies both the interlayer tunneling channels and the sublattice-symmetry-breaking hoppings. Within the same commensurate stacking topology, we fully relax the structure under hydrostatic pressures of 50 GPa and 62 GPa, obtaining interlayer distances of approximately $2.59$ and $2.51$~\AA, respectively. The bulk gap closes under compression, and isolated band crossings appear near $K$. At 50 GPa, the crossings form three symmetry-related pairs along the in-plane $C_2$ axes, as shown in Fig.~\ref{fig:Weyl_semimetal}(a). 
A representative in-plane $C_2$ axis is illustrated in Fig.~\ref{fig:unit_cell_AABB}(b) of the Supplemental Material. We calculate the Berry flux through closed surfaces enclosing the crossings and obtain quantized charges $C=\pm1$, establishing that they are Weyl points. At 62 GPa, additional pairs of Weyl points appear, as shown in Fig.~\ref{fig:Weyl_semimetal}(b). Reversing the structural chirality of the $D_6$--$D_3$ stack by changing the twist angles from $\pm21.8^\circ$ to $\mp21.8^\circ$ reverses the topological charges of the Weyl points, demonstrating a direct connection between real-space structural chirality and momentum-space topological chirality.

The pressure-induced Weyl configurations can be understood in terms of the competing tunneling channels and sublattice-symmetry breaking discussed above. Compression simultaneously changes the nonchiral hoppings $M_1$ and $M_2$, the chiral hopping $\lambda$, and the symmetry-breaking terms. The latter can gap the ideal nodal rings while leaving isolated Weyl points at symmetry-pinned zeros of the projected mass $m(\mathbf{k})$. The number and momentum-space arrangement of the Weyl points at the two pressures are consistent with remnants of the one-ring and two-ring regimes, respectively. Together with the equilibrium HOTI, these results connect the ideal phase diagram to distinct topological phases in atomistic twisted graphite.

\section{Discussion}

Our results establish the global symmetry of a large-angle interface as a distinct design degree of freedom for three-dimensional band topology. Different rotation centers generate symmetry-inequivalent intervalley-Umklapp tunneling matrices, and combining such interfaces along the stacking direction brings differently constrained chiral and nonchiral hopping channels into a single periodic structure. Their competition allows the stacking configuration to govern the electronic and topological states even at a fixed twist angle, extending the possibilities for band engineering in large-angle twisted materials.

The sublattice-symmetric model and the analysis of sublattice-symmetry-breaking perturbations provide a unified theoretical framework connecting the ideal phase diagram to the topological phases in realistic twisted graphite. The integer winding numbers of the nodal rings also determine their boundary signatures: drumhead surface states occupy the region inside the projected ring in the one-ring phase and the annular region between the projected rings in the two-ring phase [see Figs.~\ref{fig:drum_head}(c) and \ref{fig:drum_head}(d) of the Supplemental Material]. In the material realization, the reversal of Weyl charges upon reversing the structural handedness further connects real-space structural chirality to momentum-space topological chirality.

The framework also suggests opportunities beyond electronic twisted graphite. Its essential ingredients are the competition between nonchiral and chiral interlayer couplings and the symmetries that constrain their allowed forms. Photonic and acoustic lattices offer platforms in which bipartite connectivity and individual hopping amplitudes can be controlled~\cite{luTopologicalPhotonics2014,Ozawa_RMP2019,maTopologicalPhasesAcoustic2019,Acoustics_NRM}. Such control could enable exploration of the complete ideal phase diagram and the effects of deliberately introduced sublattice-symmetry breaking. 
These platforms would therefore provide complementary settings for testing how the interplay between symmetry and interlayer coupling shapes three-dimensional band topology.

\begin{acknowledgments}
\noindent{\textcolor{blue}{\it Acknowledgments}---}
We acknowledge support by the National Natural Science Foundation of China (No. 12425406), Research Grant Council of Hong Kong (AoE/P-701/20, HKU SRFS21227S05), and New Cornerstone Science Foundation.
\end{acknowledgments}


\clearpage
\onecolumngrid
\setcounter{equation}{0}
\renewcommand{\theequation}{S\arabic{equation}}
\setcounter{figure}{0}
\renewcommand{\thefigure}{S\arabic{figure}}

\renewcommand{\theHequation}{supp.\arabic{equation}}
\renewcommand{\theHfigure}{supp.\arabic{figure}}

\section*{Supplemental Material}
\setcounter{subsection}{0}
\renewcommand{\thesubsection}{S\arabic{subsection}}
\renewcommand{\theHsubsection}{supp.\arabic{subsection}}

\subsection{First-Principles Calculations}
\label{sec:sm_dft}
First-principles calculations were performed using the Vienna \textit{Ab initio} Simulation Package (VASP)~\cite{kresseEfficientIterativeSchemes1996}. The exchange--correlation interaction was described within the Perdew--Burke--Ernzerhof generalized-gradient approximation, and the projector-augmented-wave method was employed~\cite{blochlProjectorAugmentedwaveMethod1994,perdewGeneralizedGradientApproximation1996,kresseUltrasoftPseudopotentialsProjector1999}. Long-range dispersion interactions were included using the zero-damping Grimme DFT-D3 correction~\cite{grimmeConsistentAccurateInitio2010}. All calculations were non-spin-polarized, with a plane-wave energy cutoff of 600 eV.

The $D_6$--$D_3$ structures were fully optimized under hydrostatic pressures of 0, 50, and 62 GPa. During structural relaxation, a $\Gamma$-centered $9\times9\times3$ $k$-point mesh was used. Static self-consistent calculations were subsequently performed for the relaxed structures using a denser $\Gamma$-centered $12\times12\times4$ $k$-point mesh.
For the first-principles calculations of the $D_6$--$D_3$ structures under hydrostatic pressure, Wannier interpolation is used to obtain the electronic bands~\cite{pizziWannier90CommunityCode2020}.

\subsection{Slater--Koster Tight-Binding Calculations}
\label{sec:sm_sk}
For atomic Slater-Koster tight-binding calculation, the Hamiltonian is given by 
\begin{equation}
    \mathcal{H}_{\mathrm{SK}}=-\sum_{i,j}t(\bs{d}_{ij})c_i^\dagger c_j+h.c.,
\end{equation}
where $c_i^\dagger$ and $c_j$ are the creation and annihilation operators for $p_z$ orbitals on site $i$ and $j$. The hopping amplitude $t(\bs{d}_{ij})$ is determined by~\cite{slaterSimplifiedLCAOMethod1954,koshinoInterlayerInteractionGeneral2015}
\begin{equation}
\begin{split}
    -&t(\boldsymbol{d})=V_{pp\pi}\Bigg[1-\Big(\frac{\boldsymbol{d}\cdot\boldsymbol{e}_z}{d}\Big)^2\Bigg]+V_{pp\sigma}\Big(\frac{\boldsymbol{d}\cdot\boldsymbol{e}_z}{d}\Big)^2,\\
    &V_{pp\pi}=V_{pp\pi}^0\exp\Big(-\frac{d-a_0}{\delta_0}\Big),\\
    &V_{pp\sigma}=V_{pp\sigma}^0\exp\Big(-\frac{d-d_0}{\delta_0}\Big).
    \label{eq:SKTB_method}
\end{split}
\end{equation}
Here $a_0$ is the nearest-neighbor distance on monolayer graphene, $d_0$ represents the interlayer spacing, $V_{pp\pi}^0$ is the intralayer hopping between nearest-neighbor sites, and $V_{pp\sigma}^0$ denotes the energy between vertically stacked orbitals in bilayer graphene. Here, we take $a_0=1.42$ \AA, $d_0=3.35$ \AA, $\delta_0=0.68$ \AA, $V_{pp\pi}^0=-3.4$ eV, and $V_{pp\sigma}^0=0.6$ eV to fit the low-energy dispersion from DFT results.

\subsection{Momentum-Space Representation of the Effective Model}
\label{sec:sm_bloch}

Here we give the explicit Bloch Hamiltonian corresponding to Eq.~\eqref{eq:realspace} of the main text. We first group the orbitals into two bilayer sectors,
\[
\Psi_{\mathbf{k}}=
\left(\Psi_{\mathrm{AA},\mathbf{k}},\Psi_{\mathrm{BB},\mathbf{k}}\right)^T,
\]
with $\Psi_{\mathrm{AA},\mathbf{k}}=(c_{1\mathbf{k}},c_{2\mathbf{k}},c_{3\mathbf{k}},c_{4\mathbf{k}})^T$ and $\Psi_{\mathrm{BB},\mathbf{k}}=(c_{5\mathbf{k}},c_{6\mathbf{k}},c_{7\mathbf{k}},c_{8\mathbf{k}})^T$. The Bloch Hamiltonian then takes the block form
\begin{equation}
H(\mathbf{k})=
\begin{pmatrix}
H_{\mathrm{AA}}(\mathbf{k})&T_{\mathrm{AB}}(\mathbf{k})\\
T_{\mathrm{AB}}^\dagger(\mathbf{k})&H_{\mathrm{BB}}(\mathbf{k})
\end{pmatrix},
\label{eq:s_block}
\end{equation}
where
\begin{align}
H_{\mathrm{AA}}(\mathbf{k})&=
\begin{pmatrix}
0&f_{1\mathbf{k}}&g_{\mathbf{k}}^*&0\\
f_{1\mathbf{k}}^*&0&0&g_{\mathbf{k}}\\
g_{\mathbf{k}}&0&0&f_{1\mathbf{k}}\\
0&g_{\mathbf{k}}^*&f_{1\mathbf{k}}^*&0
\end{pmatrix},\notag\\[4pt]
H_{\mathrm{BB}}(\mathbf{k})&=
\begin{pmatrix}
0&f_{2\mathbf{k}}&g_{\mathbf{k}}^*&0\\
f_{2\mathbf{k}}^*&0&0&g_{\mathbf{k}}\\
g_{\mathbf{k}}&0&0&f_{2\mathbf{k}}\\
0&g_{\mathbf{k}}^*&f_{2\mathbf{k}}^*&0
\end{pmatrix}.
\label{eq:s_bilayers}
\end{align}
The AB-like coupling between the two bilayer sectors is
\begin{equation}
T_{\mathrm{AB}}(\mathbf{k})=
\begin{pmatrix}
0&0&0&h_{\mathbf{k}}\\
0&0&0&0\\
0&M_2&0&0\\
0&0&0&0
\end{pmatrix}.
\label{eq:s_tab}
\end{equation}
The matrix elements in Eqs.~\eqref{eq:s_bilayers} and \eqref{eq:s_tab} are defined using the in-plane primitive lattice vectors $\mathbf{a}_1$ and $\mathbf{a}_2$ and the out-of-plane lattice vector $\mathbf{a}_3=c\hat{\mathbf{z}}$. For the orbital convention of Fig.~\ref{fig:structure}(a),
\begin{align}
f_{1\mathbf{k}}={}&t\left[1+e^{-i\mathbf{k}\cdot\mathbf{a}_1}
+e^{-i\mathbf{k}\cdot(\mathbf{a}_1+\mathbf{a}_2)}\right],\notag\\
f_{2\mathbf{k}}={}&t\left[1+e^{i\mathbf{k}\cdot\mathbf{a}_2}
+e^{i\mathbf{k}\cdot(\mathbf{a}_1+\mathbf{a}_2)}\right],\notag\\
g_{\mathbf{k}}={}&M_1+8i\zeta\lambda
\sin\!\left(\frac{\mathbf{k}\cdot\mathbf{a}_1}{2}\right)
\sin\!\left(\frac{\mathbf{k}\cdot\mathbf{a}_2}{2}\right)\notag
\sin\!\left[\frac{\mathbf{k}\cdot(\mathbf{a}_1+\mathbf{a}_2)}{2}\right],\notag\\
h_{\mathbf{k}}={}&M_2e^{i\mathbf{k}\cdot\mathbf{a}_3}=M_2e^{ik_zc}.
\label{eq:s_elements}
\end{align}
The first term in $g_{\mathbf{k}}$ is the nonchiral AA-like interlayer hopping. Its imaginary, momentum-odd part is generated by the chiral hopping $\zeta\eta_{ij}\lambda$ shown in Fig.~\ref{fig:structure}(b). Here, $\eta_{ij}=\pm1$ specifies the bond-dependent sign pattern for a given structure, whereas $\zeta=\pm1$ labels the two opposite structural handednesses and reverses the entire chiral hopping pattern. The factor $h_{\mathbf{k}}$ describes the AB-like hopping across the boundary of the four-layer unit cell; the other AB-like connection lies within the unit cell and therefore carries no $k_z$-dependent phase.

\begin{figure}[t]
  \centering
  \includegraphics[width=0.5\columnwidth]{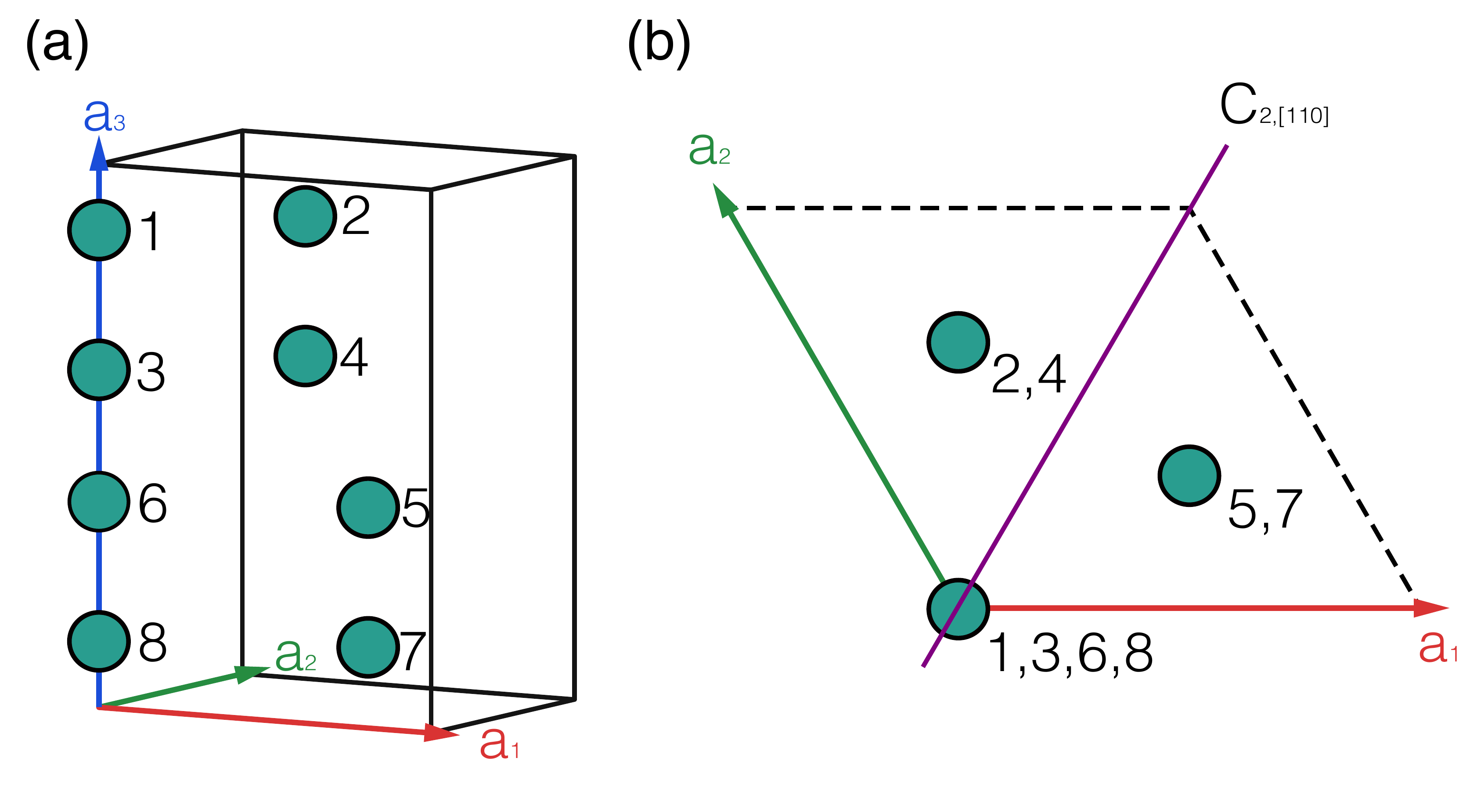}
  \caption{(a) Lattice structure of effective tight-binding model. For sublattice-symmetry analysis, the orbitals can be classified into two sublattices in \eqref{eq:sublat_indices}. (b)The top view of the lattice structure. The purple line is the rotation of axis of the in-plane $C_2$ symmetry, which pins the Weyl points in the $D_6-D_3$ stack of the main text.}
  \label{fig:unit_cell_AABB}
\end{figure}

As shown in Fig.~\ref{fig:unit_cell_AABB}, we give the diagram of the unit cell in the Eq.~\eqref{eq:realspace} of the main text. The lattice vectors are
\begin{equation}
    \begin{split}
        &\bs{a}_1=a_0 (\sqrt{3},0,0)\\
        &\bs{a}_2=a_0(-\frac{\sqrt{3}}{2},\frac{3}{2},0)\\
        &\bs{a}_3=c(0,0,1),
    \end{split}
\end{equation}
where $c$ is the out-of-plane lattice constants.

The orbitals can be classified into two effective sublattices by their indices
\begin{equation}
    \begin{split}
        &\mathrm{A\;sublattice}:(1,4,6,7)\\
        &\mathrm{B\;sublattice}:(2,3,5,8).
    \end{split}
    \label{eq:sublat_indices}
\end{equation}

For the sublattice-symmetry analysis, we further reorder the eight orbitals as
\begin{equation}
\Phi_{\mathbf{k}}=
\left(c_{1\mathbf{k}},c_{4\mathbf{k}},c_{6\mathbf{k}},c_{7\mathbf{k}};
c_{2\mathbf{k}},c_{3\mathbf{k}},c_{5\mathbf{k}},c_{8\mathbf{k}}\right)^T.
\label{eq:s_reordered}
\end{equation}
In this basis, the Bloch Hamiltonian is
\begin{equation}
H(\mathbf{k})=
\begin{pmatrix}
0&D(\mathbf{k})\\
D^\dagger(\mathbf{k})&0
\end{pmatrix},
\label{eq:s_offdiag}
\end{equation}
with
\begin{equation}
D(\mathbf{k})=
\begin{pmatrix}
f_{1\mathbf{k}}&g_{\mathbf{k}}^*&0&h_{\mathbf{k}}\\
g_{\mathbf{k}}^*&f_{1\mathbf{k}}^*&0&0\\
0&M_2&f_{2\mathbf{k}}^*&g_{\mathbf{k}}\\
0&0&g_{\mathbf{k}}&f_{2\mathbf{k}}
\end{pmatrix}.
\label{eq:s_d}
\end{equation}
This off-diagonal form makes the sublattice symmetry of the effective model explicit. Taking the determinant of Eq.~\eqref{eq:s_d} reproduces the zero-energy condition in Eq.~\eqref{eq:det} of the main text.

\subsection{Drumhead Surface States}
\label{sec:sm_drumhead}

\begin{figure}[t]
  \centering
  \includegraphics[width=0.3\columnwidth]{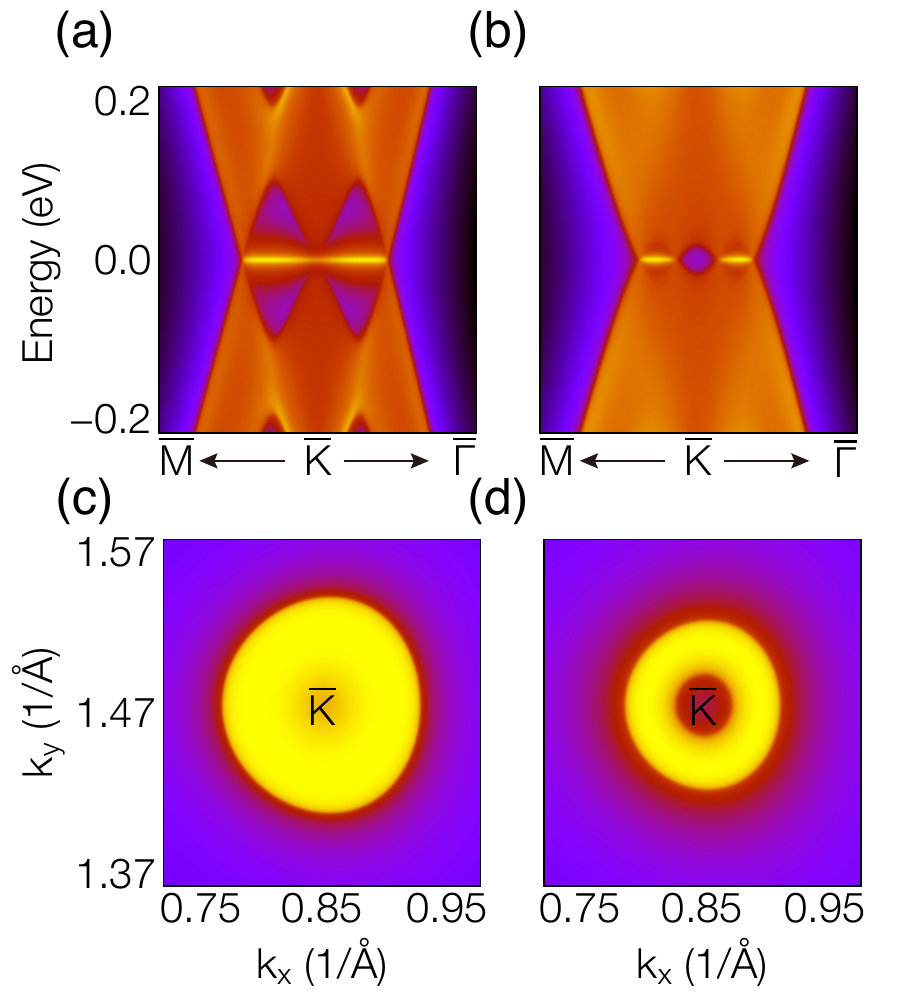}
  \caption{The (001) surface states of the 1R-NL and 2R-NL state. (a, b) Projected spectrum on the (001) surface. The drumhead surface states are located inside the projected nodal line, indicated by the yellow regions. (c, d) Projected spectrum on the (001) surface at the constant energy plane $E=0$ eV.}
  \label{fig:drum_head}
\end{figure}

Based on the toy model \eqref{eq:realspace} in the main text, the spectrum on the (001) surface near projected valley $\overline{\mathbf{K}}$ is calculated by WannierTools~\cite{wuWannierToolsOpensourceSoftware2018}. In Fig~\ref{fig:drum_head}(a) and (b), we plot the surface spectrum for 1R-NL and 2R-NL state. The parameters are $t=-2.66$ eV, $M_1=0.25$ eV and $M_2=0.27$ eV. The chiral hopping is $\lambda^\prime=0.05$ eV and $0.14$ eV for the 1R-NL and 2R-NL state, respectively. The surface states highlighted by the yellow region show the non-trivial topology of the sulattice-symmetric system. In Fig.~\ref{fig:drum_head}(c) and (d), the flat drumhead states near projected valley are depicted. It is shown that the drumhead states are located within the nodal line in the 1R-NL state and between the inner and outer nodal lines in the 2R-NL state, which is consistent with the result of topological invariant in Eq.~\ref{eq:winding}.

\end{document}